\documentclass[submission, Proceedings]{SciPost}

\hypersetup{
    colorlinks,
    linkcolor={red!50!black},
    citecolor={blue!50!black},
    urlcolor={blue!80!black}
}

\usepackage[bitstream-charter]{mathdesign}
\usepackage{wrapfig}
\usepackage{soul}

\DeclareSymbolFont{usualmathcal}{OMS}{cmsy}{m}{n}
\DeclareSymbolFontAlphabet{\mathcal}{usualmathcal}

\begin{document}
%\thispagestyle{empty}

\begin{center}{\Large \textbf{
VERITAS highlights of observations and results\\
}}\end{center}

% TODO: write the author list here. Use initials + surname format.
% Separate subsequent authors by a comma, omit comma at the end of the list.
% Mark the corresponding author with a superscript *.
\begin{center}
Patel, S. R.\textsuperscript{1*} For the VERITAS Collaboration\textsuperscript{2}
\end{center}

\begin{center}
{\textsuperscript{\bf 1}} Deutsches Elektronen-Synchrotron DESY, Platanenallee 6, D-15738 Zeuthen, Germany\\
{\textsuperscript{\bf 2}} http://veritas.sao.arizona.edu \\
* sonal.patel@desy.de
\end{center}

\begin{center}
\today
\end{center}

% For convenience during refereeing (optional),
% you can turn on line numbers by uncommenting the next line:
%\linenumbers

\definecolor{palegray}{gray}{0.95}
\begin{center}
\colorbox{palegray}{
  \begin{tabular}{rr}
  \begin{minipage}{0.1\textwidth}
    \includegraphics[width=30mm]{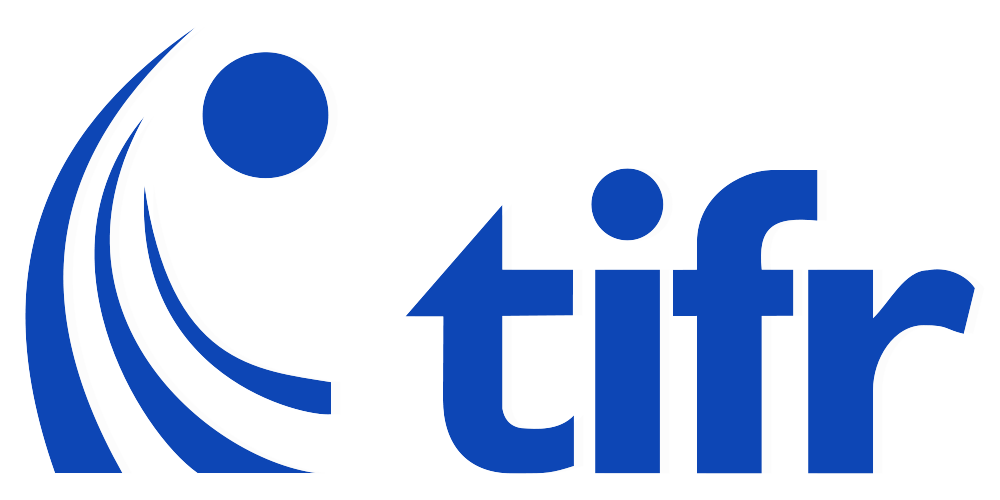}
  \end{minipage}
  &
  \begin{minipage}{0.85\textwidth}
    \begin{center}
    {\it 21st International Symposium on Very High Energy Cosmic Ray Interactions (ISVHECRI 2022)}\\
    {\it Online, 23-27 May 2022} \\
    \doi{10.21468/SciPostPhysProc.?}\\
    \end{center}
  \end{minipage}
\end{tabular}
}
\end{center}

\section*{Abstract}
{\bf
% TODO: write your abstract here.
Located in southern Arizona, VERITAS is amongst the most sensitive detectors for astrophysical very high energy (VHE; E>100 GeV) gamma rays and has been operational since April 2007. We highlight some recent results 
from VERITAS observations. These include the long-term observations of the gamma-ray binaries HESS J0632+057 
and LS I +61° 303, the observations of the Galactic Center region, and of the supernova remnant Cas~A. 
We discuss the results from a decade of multi-wavelength observations of the blazar 1ES 1215+303, the EHT 2017 
campaign on the M87 galaxy, the discovery of 3C 264 in VHE, and the observation of three 
flaring quasars. Brief highlights of the indirect dark matter searches and 
targets-of-opportunity (ToO) observations are also discussed. The ToO observations allow for rapid follow-up of multi-messenger alerts and astrophysical transients.}

\vspace{10pt}
\noindent\rule{\textwidth}{1pt}
\tableofcontents \thispagestyle{fancy}
\noindent\rule{\textwidth}{1pt}
\vspace{10pt}

\section{Introduction}
\label{sec:intro}
% TODO: write your article here.

VERITAS is one of the most sensitive ground-based $\gamma$-ray observatories. It is located at the 
Fred Lawrence Whipple Observatory in southern Arizona (31 40N, 110 57W, 1.3 km a.s.l.). The VERITAS
array has four 12-m diameter, 12-m focal length imaging atmospheric Cherenkov telescopes. 
Each telescope has a Davies-Cotton-design segmented mirror dish of 345 facets and focuses the 
Cherenkov light from particle showers onto a pixelated camera having 499 PMTs and a total field 
of view of 3.5$^\circ$. VERITAS began full operation in April 2007. Since then, it has undergone 
two major upgrades. In the first upgrade, during summer 2009, one of the four telescopes was 
relocated to a different position \cite{Perkins2009}. In the second upgrade, in summer 2012, 
PMTs (with the higher quantum efficiency and shorter time profile), and a new topological trigger 
system were installed \cite{Ziter2013}. The current configuration of the array can detect an 
object having 1$\%$ of the Crab Nebula flux in $\sim$25 hours with a gamma-ray-photon energy resolution of 15-25$\%$. 
For a 1 TeV photon, the 68$\%$ containment radius is $\le$0.1$^\circ$, with a pointing accuracy 
of <50". The details of the evolution of the performance of the VERITAS instrument with time are
discussed in \cite{Park2015}. In order to account for the changing optical throughput and detector
performance over time, signal calibration methods have recently been implemented to produce a fine-tuned
instrument response functions \cite{Adam2022}.

\section{VERITAS Observing program}

\begin{figure}[h]
\centering
\includegraphics[width=0.5\textwidth]{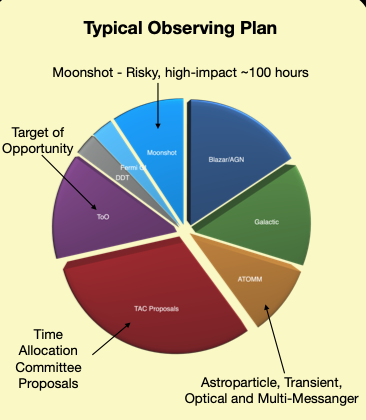}
\caption{\label{fig:observing} Typical observing plan of VERITAS}
\end{figure}

The VERITAS observing season spans from September to July each year. Each year, 
the array collects $\sim$950 h of good-weather data during dark time and $\sim$250 h 
of data during bright moon (illumination of 30-65$\%$).
The typical annual VERITAS observation plan breakdown is shown in Figure~\ref{fig:observing}.
%It can be seen that
%VERITAS observations are almost entirely proposal based %where observation proposals
%are submitted by VERITAS members and by external %scientists through a VERITAS member. 

%\begin{wrapfigure}{R}{0.5\textwidth}
%\end{wrapfigure}

\section{Galactic Science highlights}

\paragraph{$\gamma$-ray binaries - HESS J0632+057 and LS I +61$^{\circ}$ 303:} 

Based on the composition and energy output, gamma-ray binaries can be defined as systems
consisting of a compact object orbiting a star (O or Be), with periodic release of large 
amounts of non-thermal emission at energies >1 MeV \cite{Dubus2013}. The VERITAS archive 
includes one of the largest data sets for the sources of this class, which includes 
$\sim$260 h and $\sim$174 h of good quality data (after applying weather-based time cuts) 
for HESS J0632+057 and LS I +61$^{\circ}$ 303, respectively.

Located at a distance of 1.1-1.7 kpc, HESS J0632+057 consists of an unknown compact object 
orbiting a Be star (MWC 148) with a period of 321 ± 5 days \cite{Bongiorno2011}. The data from 
three major atmospheric Cherenkov telescopes: H.E.S.S. \cite{Hinton2004}, MAGIC \cite{MAGIC2016}, 
and VERITAS \cite{Park2015} were used to study the 
very high energy (VHE) $\gamma$-ray emission from this source, resulting in a total of 450 h 
over 15 years, between 2004 and 2019. The VHE $\gamma$-ray fluxes were found to be modulated with 
the orbital period of 316.7 ± 4.4(stat) ± 2.5(sys), consistent with the value obtained at X-ray
energies \cite{Adams2021}. This large data set includes dense observational coverage for several 
orbits, which reveal short-timescale and orbit-to-orbit variability.   

%\begin{figure}[!h]
%\includegraphics[width=0.45\textwidth]{HESS_J0632p57_fig1.png}
%\includegraphics[width=0.45\textwidth]{HESS_J0632p057_fig2.png}
%\caption{\label{fig:frog}Gamma-ray binaries: HESS J0632+057}
%\end{figure}

\begin{figure}[h]
\centering
\includegraphics[width=0.8\textwidth]{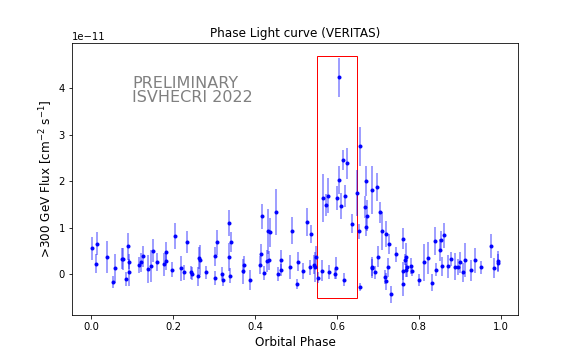}
\caption{\label{fig:lsi}Orbital phase-binned light curve of LS I +61$^\circ$ 303. The starting MJD and the orbital period used are 43366.8 days and 26.5 days, respectively.}
\end{figure}

The $\gamma$-ray binary LS I +61$^{\circ}$ 303 consists of a rapidly rotating BeVe star located 
at 2.65 ± 0.09 kpc \cite{Lindegren2021} and a compact object orbiting with a period of 26.5 days
\cite{Casares2005}. The recent detection of radio pulsations from the direction of this source by
the the Five-hundred-meter Aperture Spherical radio Telescope suggests that 
the compact object is a rotating neutron star \cite{Weng2022}. The orbital phase light curve of 
nightly-binned flux points from the analysis of $\sim$174 h of VERITAS data 
above 300 GeV is shown in Figure~\ref{fig:lsi}. The box shown in red includes the 
highest state ($\sim$30$\%$ Crab Nebula flux above 300 GeV)  of LS I +61$^{\circ}$ 303 
occurred in  October 2014, during which the flux variability on
night-timescale was observed \cite{Archambault2016}. Other than this state, 
nearly a factor of two flux difference in the orbital phase bin of (0.55-0.65) suggests 
the orbit-to-orbit variability. LS I +61$^{\circ}$ 303 is detected above 5$\sigma$ in 
most of the bins, except phase bins (0.1-0.2), (0.2-0.3), and (0.9-1.0), where the
significance is about 4$\sigma$ \cite{Kieda2022}.

\paragraph{Galactic Center Region:}

The Galactic Center (GC) region hosts numerous powerful sources and potential sites of particle
acceleration, including the supermassive black hole Sagittarius A* (Sgr A*), 
supernova remnants (SNRs), and
pulsar wind nebulae. VERITAS observes the GC at large zenith angles ($\geqslant$ 60$^\circ$),
resulting in an increased effective area of about four times greater than the effective area at 
a zenith angle of 20$^\circ$ for energies above 10 TeV, and increased in the systematic. This also raises the energy threshold
for the GC analysis to about 2 TeV. 125 h of VERITAS data resulted in a detection of Sgr A* 
at a significance of 38$\sigma$.  The differential spectrum is best fitted with a power law with an
exponential cutoff at 10.0$^{+4.0}_{-2.0}$ TeV and a spectral index of 2.12$^{+0.22}_{-0.17}$. The
analysis of the diffuse GC ridge shows a power law spectrum extending to the highest observed 
energy of $\sim$40 TeV and a hard spectral index of 2.19 ± 0.20. This supports the evidence for a
PeVatron in the GC region. A more detailed discussion of this analysis can be found in \cite{Adams2021b}.

\paragraph{Supernova Remnant - Cassiopeia A:}
SNRs are thought to be the most favorable sites for the acceleration of Galactic cosmic-rays
up to PeV energies. Among the SNRs, the young SNR are considered the best candidates to
be detected at VHE, since the highest energy cosmic rays, which are believed
to be produced early on, will not yet have escaped from the production site.  
The young ($\sim$350 years old) core-collapse SNR Cassiopeia A is a potential PeVatron candidate. VERITAS has studied this source with a deep exposure of 65 h.
The joint spectrum was produced with VERITAS data covering 200~GeV-10~TeV and 
10.8 years of $\textit{Fermi}$-Large Area Telescope data covering 0.1-500 GeV. The best fit was
obtained with the power law spectral index of 2.17 ± 0.02(stat) and cutoff energy of 
2.3 ± 0.5(stat) TeV \cite{Abeysekara2020}. Considering a one-zone model,
proton acceleration up to at least 6 TeV is required to reproduce the observed $\gamma$-ray
spectrum.

\section{Extra-galactic Science Highlights}
\label{sec:egal}
The VERITAS AGN observations are broadly conducted under four major programs, namely:
discovery program, multi-wavelength (MWL) campaigns, ToO, and cosmology.  
The highlights of the first three programs are mentioned in this section.

\paragraph{Blazar - 1ES 1215+303:}
A high-synchrotron-peaked BL Lac (HBL) 1ES 1215+303 ($z$=0.13, \cite{Truebenbach2017}) was
extensively studied in a MWL context using long-term data from radio to $\gamma$-ray energies.
a VERITAS exposure of 175.8 h was used in this study, which includes regular monitoring observations 
since December 2008. The synchrotron peak frequency from a low state to the 2017 flaring state was
observed to be shifted from infrared to soft X-ray \cite{Valverde2020}.

\paragraph{Radio Galaxies - 3C 264 and M87:}
The misaligned geometry of radio galaxies provides a unique view of the AGN's jet and super massive
back hole. Among TeV-detected radio galaxies, 3C 264 ($z=0.0217$, \cite{Smith2000}) is 
the most distant, and is only the fourth known radio galaxy at these energies. 
VERITAS has collected $\sim$57 h good quality data between February 2017 and May 2019, 
which yielded a detection with a statistical significance of 7.8$\sigma$.
The VHE $\gamma$-ray spectrum was well described by a power law with an index 
of 2.20 ± 0.27 and a flux
of $\sim$0.7$\%$ of the Crab Nebula flux above 315 GeV \cite{Archer2020}. 

During the 2017 Event Horizon Telescope MWL campaign on M87, VERITAS captured the source in its 
historically low state, but still dominating over the nearest knot, HST-1 \cite{EHT2021}. 
The most complete simultaneous MWL spectrum was reported in that study,  
which included 15 h of quality-selected VERITAS data. The analysis resulted in an
overall statistical significance of 3.8$\sigma$. The legacy data set and analysis scripts
were made available to the community through the Cyverse repository \cite{dataset}.

\paragraph{Quasars - 3C 279, PKS 1222+216, and Ton 599: }
Known flat-spectrum radio quasars (FSRQs) are generally at larger distances 
compared to BL Lacs, hence with the current generation VHE $\gamma$-rays telescopes 
their detection is difficult, both because of 
the attenuation of VHE $\gamma$-rays from these distances due to absorption by the 
extra-galactic background light and because of possible intrinsic spectral cutoffs above GeV energies
\cite{Stern2014}. Three FSRQs; 3C 279, PKS 1222+216, and Ton 599, were studied to explore the 
$\gamma$-ray variability and spectral characteristics using almost 100 h of VERITAS 
data spanning over 10 years. The location of the $\gamma$-ray emission region and the 
jet Doppler factor were 
constrained during VHE-detected flares in 2014 and 2017, for PKS 1222+216 and Ton 599, 
respectively \cite{Adams2022b}. Also, theoretical constraints on the potential production of 
PeV-scale neutrinos were placed during these VHE flares.

\section{Indirect dark matter search}
Dwarf Spheroidal galaxies (dSphs) are a favorable target class for an indirect 
dark matter (DM) search. The analysis of four dSphs (Boötes I, Draco, Segue 1, and Ursa Minor) 
with the VERITAS data taken from 2007 to 2013 was reported in \cite{Giuri2022}. 
This included a total quality-selected observation time of 476 h to search for a DM signal 
and was 
sensitive to potential signals in the $\tau^+\tau^-$ and $b\Bar{b}$ annihilation channels. No DM 
signal was detected.

\section{Multi-messenger and astrophysical transient observations}
\begin{figure}[h]
\centering
\includegraphics[width=0.9\textwidth]{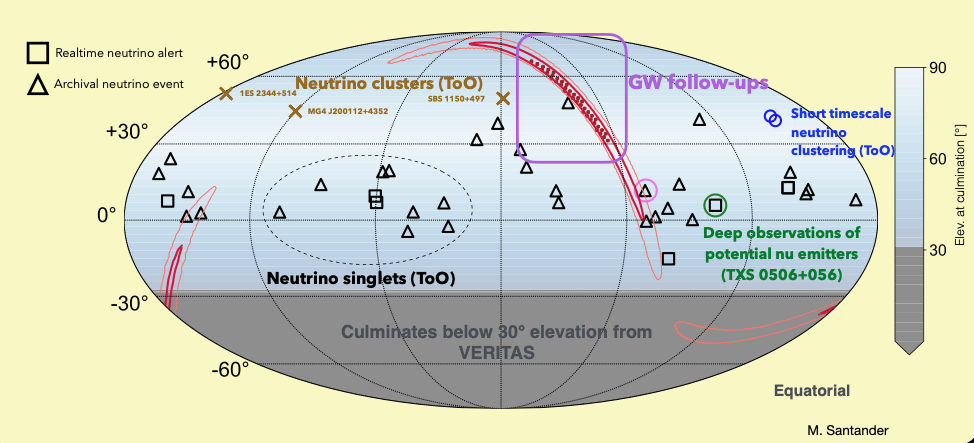}
\caption{\label{fig:mm} The sky map showing some MM triggers observed by VERITAS (Adopted from \cite{Santander2019}).}
\end{figure}

VERITAS devotes a part of its observing time to multi-messenger (MM) 
and astrophysical transient observations to search for electromagnetic counterparts 
to high energy neutrinos and gravitational waves. These observations are 
proposal driven and result into significant fraction of total VERITAS observing time. Figure~\ref{fig:mm} shows some MM triggers
observed by VERITAS. The IceCube observatory reported a well-reconstructed high energy neutrino 
event, IceCube-201114A (GCN 28887), 
on November 14, 2020, having an estimated energy of $\sim$214 TeV. It is spatially
coincident with the high-energy-peaked object, NVSS J065844+063711 \cite{Menezes2022}.
During November 15-19, 2020, VERITAS observed NVSS J065844+063711,
and a differential upper limit was reported in \cite{Menezes2022} using 7 h quality-selected data.

\section{Summary and Conclusions}
VERITAS has a strong and multi-faceted science program in the $\gamma$-ray band. 
Selected science results are highlighted, covering results from Galactic, extra-galactic,
fundamental physics, and multi-messenger observations. VERITAS is operating extremely 
well and continues to provide high quality VHE $\gamma$-ray data and scientific results 
to the community. VERITAS has been recommended for the next cycle of NSF operations funding through 2025.

\section*{Acknowledgements}
This research is supported by grants from the U.S. Department of Energy Office of Science, the U.S. National Science Foundation and the Smithsonian Institution, by NSERC in Canada, and by the Helmholtz Association in Germany. This research used resources provided by the Open Science Grid, which is supported by the National Science Foundation and the U.S. Department of Energy's Office of Science, and resources of the National Energy Research Scientific Computing Center (NERSC), a U.S. Department of Energy Office of Science User Facility operated under Contract No. DE-AC02-05CH11231. We acknowledge the excellent work of the technical support staff at the Fred Lawrence Whipple Observatory and at the collaborating institutions in the construction and operation of the instrument.

%Acknowledgements should follow immediately after the conclusion.
% TODO: include author contributions
%\paragraph{Author contributions}

% TODO: include funding information
%\paragraph{Funding information}
%Authors are required to provide funding information, including relevant agencies and grant numbers with linked author's initials. Correctly-provided data will be linked to funders listed in the \href{https://www.crossref.org/services/funder-registry/}{\sf Fundref registry}.

\bibliographystyle{SciPost_bibstyle}
%\bibliography{SciPost_Example_BiBTeX_File.bib}
\bibliography{ISVHECRI.bib}

\begin{thebibliography}{10}
\providecommand{\url}[1]{\texttt{#1}}
\providecommand{\urlprefix}{URL }
\expandafter\ifx\csname urlstyle\endcsname\relax
  \providecommand{\doi}[1]{doi:\discretionary{}{}{}#1}\else
  \providecommand{\doi}{doi:\discretionary{}{}{}\begingroup
  \urlstyle{rm}\Url}\fi
\providecommand{\eprint}[2][]{\url{#2}}

\bibitem{Perkins2009}
J.~S. {Perkins}, G.~{Maier} and {The VERITAS Collaboration},
\newblock \emph{{VERITAS Telescope 1 Relocation: Details and Improvements}},
\newblock arXiv e-prints arXiv:0912.3841 (2009),
\newblock \eprint{0912.3841}.

\bibitem{Ziter2013}
B.~{Zitzer} and {VERITAS Collaboration},
\newblock \emph{{The VERITAS Upgraded Telescope-Level Trigger Systems:
  Technical Details and Performance Characterization}},
\newblock In \emph{International Cosmic Ray Conference}, vol.~33 of
  \emph{International Cosmic Ray Conference}, p. 3076 (2013),
  \eprint{1307.8360}.

\bibitem{Park2015}
N.~{Park} and {VERITAS Collaboration},
\newblock \emph{{Performance of the VERITAS experiment}},
\newblock In \emph{34th International Cosmic Ray Conference (ICRC2015)},
  vol.~34 of \emph{International Cosmic Ray Conference}, p. 771 (2015),
  \eprint{1508.07070}.

\bibitem{Adam2022}
C.~B. {Adams}, W.~{Benbow}, A.~{Brill}, J.~H. {Buckley}, J.~L. {Christiansen},
  A.~{Falcone}, Q.~{Feng}, J.~P. {Finley}, G.~M. {Foote}, L.~{Fortson},
  A.~{Furniss}, C.~{Giuri} \emph{et~al.},
\newblock \emph{{The throughput calibration of the VERITAS telescopes}},
\newblock Astronomy \& Astrophysics \textbf{658}, A83 (2022),
\newblock \doi{10.1051/0004-6361/202142275},
\newblock \eprint{2111.04676}.

\bibitem{Dubus2013}
G.~{Dubus},
\newblock \emph{{Gamma-ray binaries and related systems}},
\newblock The Astronomy and Astrophysics Review \textbf{21}, 64 (2013),
\newblock \doi{10.1007/s00159-013-0064-5},
\newblock \eprint{1307.7083}.

\bibitem{Bongiorno2011}
S.~D. {Bongiorno}, A.~D. {Falcone}, M.~{Stroh}, J.~{Holder}, J.~L. {Skilton},
  J.~A. {Hinton}, N.~{Gehrels} and J.~{Grube},
\newblock \emph{{A New TeV Binary: The Discovery of an Orbital Period in HESS
  J0632+057}},
\newblock The Astrophysical Journal Letters \textbf{737}(1), L11 (2011),
\newblock \doi{10.1088/2041-8205/737/1/L11},
\newblock \eprint{1104.4519}.

\bibitem{Hinton2004}
J.~A. {Hinton} and {HESS Collaboration},
\newblock \emph{{The status of the HESS project}},
\newblock New Astronomy Reviews \textbf{48}(5-6), 331 (2004),
\newblock \doi{10.1016/j.newar.2003.12.004},
\newblock \eprint{astro-ph/0403052}.

\bibitem{MAGIC2016}
J.~{Aleksi{\'c}}, S.~{Ansoldi}, L.~A. {Antonelli}, P.~{Antoranz}, A.~{Babic},
  P.~{Bangale}, M.~{Barcel{\'o}}, J.~A. {Barrio}, J.~{Becerra Gonz{\'a}lez},
  W.~{Bednarek}, E.~{Bernardini}, B.~{Biasuzzi} \emph{et~al.},
\newblock \emph{{The major upgrade of the MAGIC telescopes, Part II: A
  performance study using observations of the Crab Nebula}},
\newblock Astroparticle Physics \textbf{72}, 76 (2016),
\newblock \doi{10.1016/j.astropartphys.2015.02.005},
\newblock \eprint{1409.5594}.

\bibitem{Adams2021}
C.~B. {Adams}, W.~{Benbow}, A.~{Brill}, J.~H. {Buckley}, M.~{Capasso}, A.~J.
  {Chromey}, M.~{Errando}, A.~{Falcone}, K.~{A Farrell}, Q.~{Feng}, J.~P.
  {Finley}, G.~{M Foote} \emph{et~al.},
\newblock \emph{{Observation of the Gamma-Ray Binary HESS J0632+057 with the
  H.E.S.S., MAGIC, and VERITAS Telescopes}},
\newblock The Astrophysical Journal \textbf{923}(2), 241 (2021),
\newblock \doi{10.3847/1538-4357/ac29b7},
\newblock \eprint{2109.11894}.

\bibitem{Lindegren2021}
L.~{Lindegren}, S.~A. {Klioner}, J.~{Hern{\'a}ndez}, A.~{Bombrun},
  M.~{Ramos-Lerate}, H.~{Steidelm{\"u}ller}, U.~{Bastian}, M.~{Biermann},
  A.~{de Torres}, E.~{Gerlach}, R.~{Geyer}, T.~{Hilger} \emph{et~al.},
\newblock \emph{{Gaia Early Data Release 3. The astrometric solution}},
\newblock Astronomy \& Astrophysics \textbf{649}, A2 (2021),
\newblock \doi{10.1051/0004-6361/202039709},
\newblock \eprint{2012.03380}.

\bibitem{Casares2005}
J.~{Casares}, I.~{Ribas}, J.~M. {Paredes}, J.~{Mart{\'\i}} and C.~{Allende
  Prieto},
\newblock \emph{{Orbital parameters of the microquasar LS I +61 303}},
\newblock Monthly Notices of the Royal Astronomical Society \textbf{360}(3),
  1105 (2005),
\newblock \doi{10.1111/j.1365-2966.2005.09106.x},
\newblock \eprint{astro-ph/0504332}.

\bibitem{Weng2022}
S.-S. {Weng}, L.~{Qian}, B.-J. {Wang}, D.~F. {Torres}, A.~{Papitto},
  P.~{Jiang}, R.~{Xu}, J.~{Li}, J.-Z. {Yan}, Q.-Z. {Liu}, M.-Y. {Ge} and Q.-R.
  {Yuan},
\newblock \emph{{Radio pulsations from a neutron star within the gamma-ray
  binary LS I +61{\textdegree} 303}},
\newblock Nature Astronomy \textbf{6}, 698 (2022),
\newblock \doi{10.1038/s41550-022-01630-1},
\newblock \eprint{2203.09423}.

\bibitem{Archambault2016}
S.~{Archambault}, A.~{Archer}, T.~{Aune}, A.~{Barnacka}, W.~{Benbow},
  R.~{Bird}, M.~{Buchovecky}, J.~H. {Buckley}, V.~{Bugaev}, K.~{Byrum}, J.~V.
  {Cardenzana}, M.~{Cerruti} \emph{et~al.},
\newblock \emph{{Exceptionally Bright TeV Flares from the Binary LS I +61
  303}},
\newblock The Astrophysical Journal Letters \textbf{817}(1), L7 (2016),
\newblock \doi{10.3847/2041-8205/817/1/L7},
\newblock \eprint{1601.01812}.

\bibitem{Kieda2022}
D.~{Kieda} and {et al.},
\newblock \emph{{Very High Energy Gamma-ray Emission from the Binary System LS
  I +61 303}},
\newblock In \emph{37th International Cosmic Ray Conference. 12-23 July 2021.
  Berlin}, p. 832 (2022).

\bibitem{Adams2021b}
C.~B. {Adams}, W.~{Benbow}, A.~{Brill}, R.~{Brose}, M.~{Buchovecky},
  M.~{Capasso}, J.~L. {Christiansen}, A.~J. {Chromey}, M.~K. {Daniel},
  M.~{Errando}, A.~{Falcone}, Q.~{Feng} \emph{et~al.},
\newblock \emph{{VERITAS Observations of the Galactic Center Region at
  Multi-TeV Gamma-Ray Energies}},
\newblock The Astrophysical Journal \textbf{913}(2), 115 (2021),
\newblock \doi{10.3847/1538-4357/abf926},
\newblock \eprint{2104.12735}.

\bibitem{Abeysekara2020}
A.~U. {Abeysekara}, A.~{Archer}, W.~{Benbow}, R.~{Bird}, R.~{Brose},
  M.~{Buchovecky}, J.~H. {Buckley}, A.~J. {Chromey}, W.~{Cui}, M.~K. {Daniel},
  S.~{Das}, V.~V. {Dwarkadas} \emph{et~al.},
\newblock \emph{{Evidence for Proton Acceleration up to TeV Energies Based on
  VERITAS and Fermi-LAT Observations of the Cas A SNR}},
\newblock The Astrophysical Journal \textbf{894}(1), 51 (2020),
\newblock \doi{10.3847/1538-4357/ab8310},
\newblock \eprint{2003.13615}.

\bibitem{Truebenbach2017}
A.~E. {Truebenbach} and J.~{Darling},
\newblock \emph{{The VLBA Extragalactic Proper Motion Catalog and a Measurement
  of the Secular Aberration Drift}},
\newblock The Astrophysical Journal Supplement Series \textbf{233}(1), 3
  (2017),
\newblock \doi{10.3847/1538-4365/aa9026}.

\bibitem{Valverde2020}
J.~{Valverde}, D.~{Horan}, D.~{Bernard}, S.~{Fegan}, {Fermi-LAT Collaboration},
  A.~U. {Abeysekara}, A.~{Archer}, W.~{Benbow}, R.~{Bird}, A.~{Brill},
  R.~{Brose}, M.~{Buchovecky} \emph{et~al.},
\newblock \emph{{A Decade of Multiwavelength Observations of the TeV Blazar 1ES
  1215+303: Extreme Shift of the Synchrotron Peak Frequency and Long-term
  Optical-Gamma-Ray Flux Increase}},
\newblock The Astrophysical Journal \textbf{891}(2), 170 (2020),
\newblock \doi{10.3847/1538-4357/ab765d},
\newblock \eprint{2002.04119}.

\bibitem{Smith2000}
R.~J. {Smith}, J.~R. {Lucey}, M.~J. {Hudson}, D.~J. {Schlegel} and R.~L.
  {Davies},
\newblock \emph{{Streaming motions of galaxy clusters within 12000kms$^{-1}$ -
  I. New spectroscopic data}},
\newblock Monthly Notices of the Royal Astronomical Society \textbf{313}(3),
  469 (2000),
\newblock \doi{10.1046/j.1365-8711.2000.03251.x}.

\bibitem{Archer2020}
A.~{Archer}, W.~{Benbow}, R.~{Bird}, A.~{Brill}, M.~{Buchovecky}, J.~H.
  {Buckley}, M.~T. {Carini}, J.~L. {Christiansen}, A.~J. {Chromey}, M.~K.
  {Daniel}, M.~{Errando}, A.~{Falcone} \emph{et~al.},
\newblock \emph{{VERITAS Discovery of VHE Emission from the Radio Galaxy 3C
  264: A Multiwavelength Study}},
\newblock The Astrophysical Journal \textbf{896}(1), 41 (2020),
\newblock \doi{10.3847/1538-4357/ab910e},
\newblock \eprint{2005.03110}.

\bibitem{EHT2021}
{EHT MWL Science Working Group}, J.~C. {Algaba}, J.~{Anczarski}, K.~{Asada},
  M.~{Balokovi{\'c}}, S.~{Chandra}, Y.~Z. {Cui}, A.~D. {Falcone},
  M.~{Giroletti}, C.~{Goddi}, K.~{Hada}, D.~{Haggard} \emph{et~al.},
\newblock \emph{{Broadband Multi-wavelength Properties of M87 during the 2017
  Event Horizon Telescope Campaign}},
\newblock The Astrophysical Journal Letters \textbf{911}(1), L11 (2021),
\newblock \doi{10.3847/2041-8213/abef71},
\newblock \eprint{2104.06855}.

\bibitem{dataset}
T.~E. M. S.~W. Group,
\newblock \emph{Multi-wavelength observations of m87 during the 2017 event
  horizon telescope campaign},
\newblock \doi{10.25739/mhh2-cw46} (2021).

\bibitem{Stern2014}
B.~E. {Stern} and J.~{Poutanen},
\newblock \emph{{The Mystery of Spectral Breaks: Lyman Continuum Absorption by
  Photon-Photon Pair Production in the Fermi GeV Spectra of Bright Blazars}},
\newblock The Astrophysical Journal \textbf{794}(1), 8 (2014),
\newblock \doi{10.1088/0004-637X/794/1/8},
\newblock \eprint{1408.0793}.

\bibitem{Adams2022b}
C.~B. {Adams}, J.~{Batshoun}, W.~{Benbow}, A.~{Brill}, J.~H. {Buckley},
  M.~{Capasso}, B.~{Cavins}, J.~L. {Christiansen}, P.~{Coppi}, M.~{Errando},
  K.~A. {Farrell}, Q.~{Feng} \emph{et~al.},
\newblock \emph{{Variability and Spectral Characteristics of Three Flaring
  Gamma-Ray Quasars Observed by VERITAS and Fermi-LAT}},
\newblock The Astrophysical Journal \textbf{924}(2), 95 (2022),
\newblock \doi{10.3847/1538-4357/ac32bd},
\newblock \eprint{2110.13181}.

\bibitem{Giuri2022}
C.~{Giuri} and {VERITAS Collaboration},
\newblock \emph{{VERITAS dark matter search in dwarf Spheroidal galaxies: an
  extended source analysis}},
\newblock In \emph{37th International Cosmic Ray Conference. 12-23 July 2021.
  Berlin}, p. 515 (2022), \eprint{2108.06083}.

\bibitem{Menezes2022}
R.~{de Menezes}, S.~{Buson}, S.~{Garrappa}, A.~{Gokus}, M.~{Kadler},
  T.~{Cheung}, M.~{Giroletti}, M.~{Ajello}, F.~{Massaro}, H.~{Pe{\~n}a-Herazo},
  F.~{Sch{\"u}ssler}, E.~{Bernardini} \emph{et~al.},
\newblock \emph{{Multi-Messenger observations of the Fermi-LAT blazar 4FGL
  J0658.6+0636 consistent with an IceCube high-energy neutrino}},
\newblock In \emph{37th International Cosmic Ray Conference. 12-23 July 2021.
  Berlin}, p. 955 (2022).

\bibitem{Santander2019}
M.~{Santander},
\newblock \emph{{Recent results from the VERITAS multi-messenger program}},
\newblock In \emph{36th International Cosmic Ray Conference (ICRC2019)},
  vol.~36 of \emph{International Cosmic Ray Conference}, p. 782 (2019),
  \eprint{1909.05228}.

\end{thebibliography}

\nolinenumbers

\end{document}